# Modeling the sense of presence of remote participants in hybrid communication and its application to the design of avatar robot behavior

Takuma Miyaguchi[1] and Hideyoshi Yanagisawa[1]*

[1] Graduate School of Engineering, The University of Tokyo, Tokyo, Japan

* Corresponding author
E-mail: hide@mech.t.u-tokyo.ac.jp (HY)

# Abstract

Hybrid face-to-face and online communication have the problem that only the local participants are stimulated and the remote participants are excluded or the local participants forget about the remote participants. These problems were considered as the lack of sense of the presence of remote participants. As the first quantitative approach to the presence, to the best of our knowledge, we formulated the sense of the presence of a remote participant in hybrid communication using a Bayesian framework. We also applied the knowledge gained from the simulation with the Bayesian model to the avatar robot's intervention behavior and encouraged the local participants to speak by intervening in the remote participant's behavior using an avatar robot. We hypothesized that beliefs about the remote participant's mental states would be a factor in the remote participant's presence. We formulated presence as model evidence for other's mental state. We then modeled the influence of the avatar robot's behavior on the local participants' statements using an active inference framework that included the presence of a remote participant as a latent variable. Through simulations with the model, it was possible to predict that the gaze behavior of the avatar robot would encourage local participants to speak. Based on the simulation results, we designed the gaze behavior of an avatar robot. Finally, we examined the effectiveness of the designed gaze behavior of the avatar robot. The gaze behavior expressed more of the remote participant's attention and interest in local participants, but local participants expressed fewer opinions in the meeting tasks. The results suggest that gaze behavior increased the presence of the remote participant and discouraged the local participant from speaking in the context of the experimental task. We believe that presence has a sufficiently large influence on whether participants want to express an opinion. It is worth investigating the influence of presence and its control methods using Bayesian models.

## Introduction

In recent years, many conferences and lectures have been conducted in a hybrid format combining face-to-face and online communication. Problems with hybrid communication have been reported [1]. In particular, the problem is that only local participants are stimulated and the remote participants are excluded and the local participants forget the remote participants. Several studies have attributed the problems encountered in hybrid communication to a lack of remote participants and have attempted to increase their presence. For example, Yankelovich et al. developed a device that combines a remotely controllable display and a high-quality speaker to increase the social presence of remote participants [1]. Van Dijk et al. proposed a 3D interface that can naturally reflect gaze direction in images by arranging the placement of cameras and images and investigating the effects on the perception of social presence [2]. However, these studies are limited to qualitative discussions, and to our knowledge, a quantitative approach to presence using mathematical models has not yet been proposed.

The mathematical model allows us to comprehensively predict the effects of changes in latent variables on the presence and the effects of presence. These predictions may exceed human intuition. In addition, we can research along the cycle: first, we predict the effect of changes in the set of latent variables on presence through simulations and then discover new latent variables through experiments and update the model. Thus, we believe that proposing a mathematical model for presence will enable a more effective approach to problems related to presence.

Ogasawara et al. formulated a sense of existence for other avatars in VR using a Bayesian perceptual model [3]. They modeled the sense of existence as the model evidence for the spatiotemporal localization of the avatar. In short, they physically modeled the senses of the avatar. However, we have formulated the presence of remote participants, which causes problems in hybrid communication. By exploring the causes of these problems, we set up a hypothesis about the factors of presence.

In addition, using a mathematical model of presence, we propose an avatar robot behavior that promotes the statements of local participants. In groups with different status levels, low-status members tend to be reluctant to express their opinions about interpersonal risks. We used an avatar robot to intervene in the behavior of high-status members in an attempt to solve this problem.

This study has two objectives. First, the presence of a remote participant in hybrid communication is formulated. Second, we apply findings from the simulation with a Bayesian model to the intervening behavior of the avatar robot and promote the local participant's statement by intervening with the avatar robot in the remote participant's behavior.

In the *mathematical modeling of the sense of the presence of a remote person,* we first explain the conversational experiment with an avatar robot and the hypothesis for the presence factor. We then introduce active inference and explain the mathematical modeling of the remote participants' presence.

In *designing the gaze behavior of the avatar robot that encourages the expression of opinions by local participants in the discussion,* we explain the mathematical modeling and simulations for designing the avatar robot's behavior that promotes local participants' statements. The focus was on statements made in groups with different social status levels. The model considers the presence of a remote participant as a latent variable. In addition, an avatar robot's gaze behavior based on the simulation findings is proposed.

In *Experiment: Effect of the intervention in the gaze behavior of the avatar robot on the statements of local participants,* we describe an experiment to investigate the effectiveness of the designed gaze behavior. We recreated a situation in which low-status members were reluctant to express their opinions with groups with different social status levels.

# Methods

## Significance of the use of the avatar robot

In this study, an avatar robot as a device for remote participation is used in hybrid communication. The avatar robot is a prototype developed by Sony Group Corporation, with whom we are conducting joint research. Fig 1 shows an overview of an avatar robot. The avatar robot plays the voice of the remote party through the speaker. The avatar robot also mirrors the remote participant's gaze, blink, and head rotation.

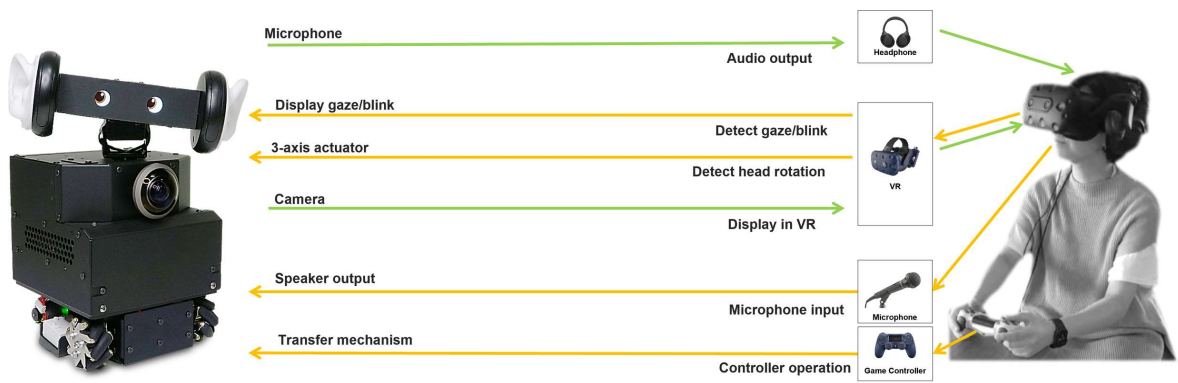

**Fig 1. Avatar robot used in this study.**

These two features are important for the use of avatar robots. First, the avatar robot accurately reproduced the gaze direction of the remote participant. In hybrid systems that use a plane display, it is difficult to determine the gaze direction of a remote participant. It has been reported that the transmission of gaze direction is more accurate when a robot's three-dimensional head rotates than when a remote participant's face appears on a display [4]. The avatar robot rotates its head along with the head rotation of the remote participant. We believe that the avatar robot can accurately reproduce the gaze direction.

Second, the avatar robot can intervene in the transmission of a remote participant's nonverbal behavior (i.e., head rotation, gazing, and blinking). The avatar robot not only reflects the behavior of the remote participant but also edits, transforms, or automates it. Non-verbal behaviors express emotions [5] and cognitive states [6]. We have attempted to control presence by intervening in non-verbal behaviors and controlling the transmission of emotions and cognitive states.

## Mathematical modeling of the sense of the presence of a remote person

### Conversational experiment with an avatar robot to formulate hypotheses about the definition and factors of presence.

We conducted a conversational experiment with an avatar robot as a first step in the mathematical modeling of presence. The purpose was to formulate a hypothesis about the definition and factors of the presence of remote participants.

One remote participant and three local participants took part in a conversation task. Fig 2 shows the arrangement of participants in the conversational experiment. Two conversation formats (presentation and discussion) were set as tasks. The reason for this is that it is necessary to observe or experience different conversation scenes (i.e., where one participant talks to others, where there are free alternations, and where one participant addresses another).

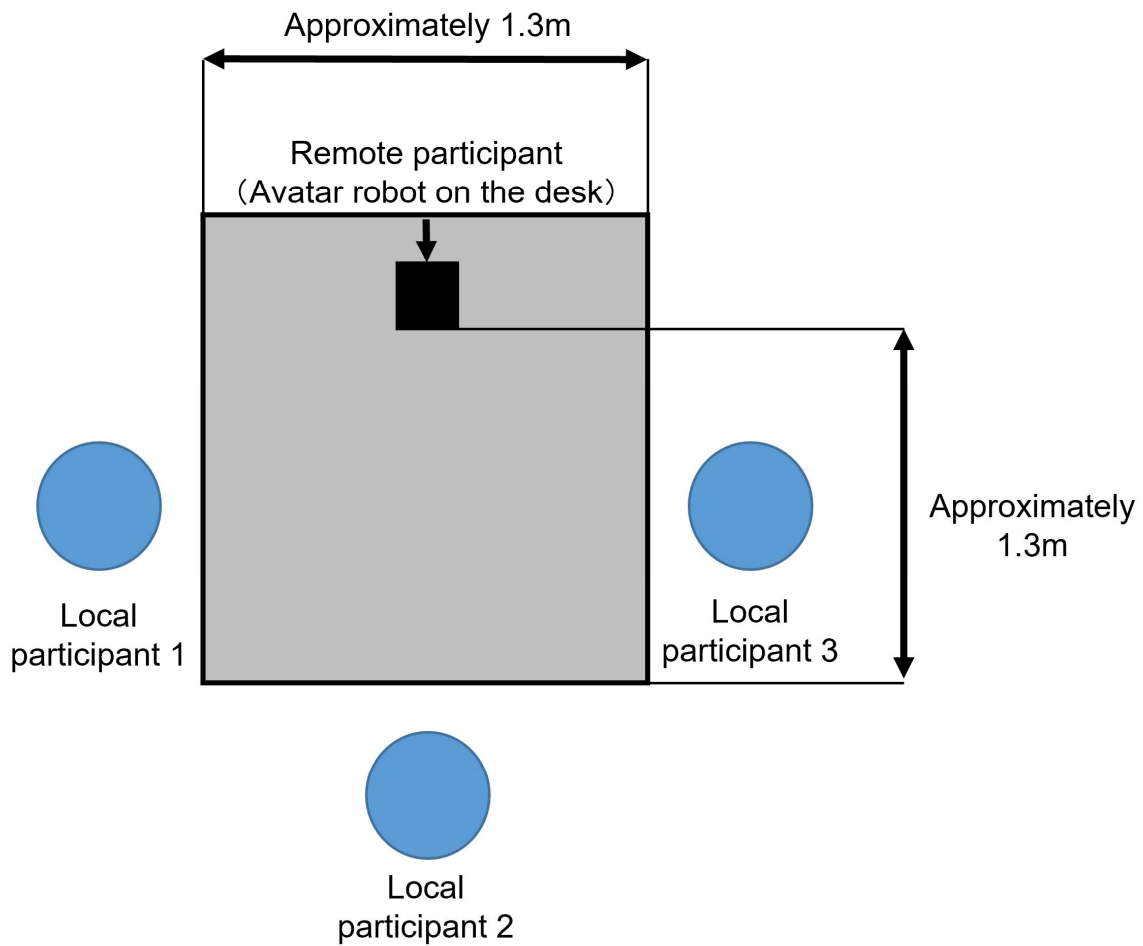

Fig 2. Arrangement of participants in the conversation experiment.

The presence being investigated is attributed to problems in hybrid communication, such as the problem that only local participants are stimulated, that remote participants are excluded, or that the local participants forget the remote participants. Therefore, we discuss the reasons why only local participants become excited, why remote participants are excluded, or why they forget about remote participants. The following reasons were cited as causes:
- It is difficult for remote participants to take turns during active alternation because only remote participants experience a delay.

- It is difficult for local participants to understand the mental states of remote participants, such as their thoughts and cognitive states.
- It is difficult for local participants to allow remote participants to take turns because they cannot predict the remote participants' responses.

Of these three points, we focused on the bottom two, which pertain to the recognition of remote participants. The avatar robot can display a remote participant's gaze, blink, and head rotation, but it cannot display other behaviors, such as facial expressions and hand gestures. A hybrid system that shows a remote participant on a display can represent the remote participant's facial expressions and hand gestures, but it is difficult to know where the remote participant is looking. In any hybrid system, the cues to the mental states of the remote participants were less than in face-to-face communication. Therefore, it is difficult to determine the mental state of a remote participant. In addition, it is difficult to predict remote participants' responses when they are unaware of their mental states. The risk of communication errors, such as speaking to a remote participant who is not ready to respond and the difficulty of speaking in accordance with the remote participant's thoughts, seems to be the cause of the difficulties in talking to remote participants.

We hypothesized that the difficulty in detecting the mental state of a remote participant was related to problems in hybrid communication and the presence of remote participants. We also hypothesized that beliefs about remote participants' mental states would be a factor in remote participants' presence. In this study, we used the active inference framework [7] to mathematically capture beliefs about a remote participant's mental state.

## Active inference

In this section, the active inference method is described, which is used to model the sense of presence. A detailed explanation can be found in [7].

Active inference is a mathematical framework based on the idea that perception and learning can be understood as minimizing variational free energy (VFE) and action selection and planning. Decision-making can be understood as minimizing the expected free energy (EFE) [7]. To explain active inference, Bayesian inference is discussed first.

Bayesian inference uses Bayes' theorem to infer the hidden states of the external world from the observations generated by these hidden states. The term "hidden" here means that the state itself cannot be observed directly but can only be inferred from observation (sensory stimuli). Eq. (1) shows the Bayes' theorem.

$$p(s|o) = \frac{p(o|s)p(s)}{p(o)} \quad (1)$$

The hidden state is denoted by s, and the observation is denoted by o. The term $p(s|o)$ is the posterior encoding belief regarding the state obtained by inference. The term $p(o|s)$ is the likelihood of encoding knowledge about the mapping between the observation and state. The term $p(s)$ is the prior encoding prediction of the state before the observation. The term $p(o)$ is the model evidence encoding the plausibility of the generative model. Bayes' theorem states that the likelihood and prior are combined to derive the posterior.

Bayes' theorem, however, cannot be computed directly for all but the simplest distributions. This is because calculating the model evidence requires summing the probabilities of the observations under all possible states of the generative model. Eq. (2) shows the formula for model evidence:

$$p(o) = \sum_s p(o,s) = \sum_s p(o|s)p(s) \quad (2)$$

Therefore, variational inference was performed to approximate the posterior. An approximate posterior $q(s)$ is introduced. The VFE was then calculated and expressed as the Kullback-Leibler divergence (KL divergence) between $q(s)$ and the generative model $p(o,s)$. KL divergence corresponds to a measure that calculates the dissimilarity between two distributions. Eq. (3) shows the VFE.

$$F = \sum_s q(s) \ln \frac{q(s)}{p(o,s)} \quad (3)$$

Variationally update $q(s)$. If $q(s)$ is found to minimize the VFE, then approximate the true posterior $p(s|o)$ at that point.

This explains the perceptual mechanism in active inferences. Next, action selection is discussed.

In action selection, the brain also makes decisions that provide future observations to minimize the VFE. However, future outcomes have not yet been observed. Therefore, the EFE (i.e., the expected value of the VFE) was calculated and minimized. The expected free energy of the action policy π is expressed in Eq. (4) [7]:

$$G_\pi = \sum_{s,o} q(o,s|\pi) \ln \frac{q(s|\pi)}{p(o,s|\pi)} \quad (4)$$

$$= \sum_{s,o} q(o,s|\pi) \ln \frac{q(s|\pi)}{p(s|o,\pi)} - \sum_{o} q(o|\pi) \ln p(o|\pi)$$

$$\approx \sum_{s,o} q(o,s|\pi) \ln \frac{q(s|\pi)}{q(s|o,\pi)} - \sum_{o} q(o|\pi) \ln p(o|C)$$

Two substitutions are made in the third line of Eq. (4). First, the approximate posterior $q(o,s|\pi)$ replaces the true posterior $p(s|o,\pi)$. Second, $p(o|C)$ replaces $p(o|\pi)$ in the second term. C is the prior preference, and $p(o|C)$ encodes the observation that the agent prefers based on the prior preference.

## Mathematical modeling of presence using an active inference framework

The sense of the presence of the remote participant in hybrid communication is modeled using active inference.

Ogasawara et al. modeled the sense of the existence of another person's avatar in VR as the magnitude of model evidence or the smallness of surprise when perceiving stimuli related to the spatiotemporal localization of the avatar [3]. We hypothesized that beliefs about the mental states of remote participants would be a factor in the presence of the remote participants in the conversation experiment. The sense of presence by considering beliefs about a remote participant's mental state was modeled by using the model proposed by Ogasawara et al.

The sense of the presence of remote participants in hybrid communication was modeled as the model evidence of the mental states of remote participants. The large amount of model evidence for the mental state of a remote participant indicates its clarity. Therefore, the model evidence is associated with the presence of remote participants. With this model, the presence of remote participants can be expressed as latent variables in an active inference framework.

## Designing the gaze behavior of the avatar robot that encourages the expression of opinions by local participants in the discussion

In this section, the behavior of the avatar robot is designed that encourages the expression of opinions by local participants in the discussion and solves the problem of expressing opinions in groups with different social status levels. For this purpose, the influence of the gaze behavior of a high-status member on the statements of the low-status members was modeled, including the presence modeled in the active inference framework. Then, simulations using the model were performed, and insights were obtained for designing the avatar robot's behavior.

**Problems in expressing opinions in groups with different levels of social status**

In group discussions, sharing information and exchanging opinions are important because they enhance the performance of the group. Sharing information can allow groups with divergent knowledge or recognition levels to pool and summarize all information [8]. It is reported that that disagreement in groups is associated with increased creativity [9] and better decision making [10].

However, voicing opinions to share information or express dissenting views is often associated with interpersonal risks. Speaking out against the prevailing opinion in the group can lead to social exclusion from the group [11]. Expressing opinions can elicit critical feedback and create an impression of incompetence in others. If there are people in the group who can influence their promotion or salary, expressing their opinions can be a more concrete and risky action [12]. Therefore, in groups with different status levels, the problem arises that low-status members hesitate to express their opinions because they fear interpersonal risks.

**Mathematical modeling of the influence of the gaze behavior of a high-status member on the statements of low-status members**

The characteristics and behaviors of high-status members influence the statements of others. Whether the team leader is cooperative or coaching-oriented has been reported to be a factor predicting a team's psychological safety [12]. Psychological safety is the shared belief that a team is safe to take risks in interpersonal relationships and is related to team members' learning behaviors [12]. Shim et al. found that differences in group members' influence decreased when a group leader directed more gaze toward a low-status member. [13].

We promoted the statements of low-status members by intervening in the non-verbal behavior of high-status members using the avatar robot. To design the behavior of an avatar robot to promote the statements of low-status members, the influence of the behavior of a high-status member on the statements of low-status members was modeled using an active inference framework.

Fig 3 shows the proposed Bayesian model of the influence of a high-status member's behavior on the statements of low-status members. s is the hidden state; o is the observation; A is likelihood mapping between states and observations (observation model); B is the transition matrix; C is the prior preferences; and D is the initial state priors. The model was divided into three fields: Level 1-1, Level 2, and Level 1-2. In all fields, the agent of inference is a low-status member. The fields are described as follows:

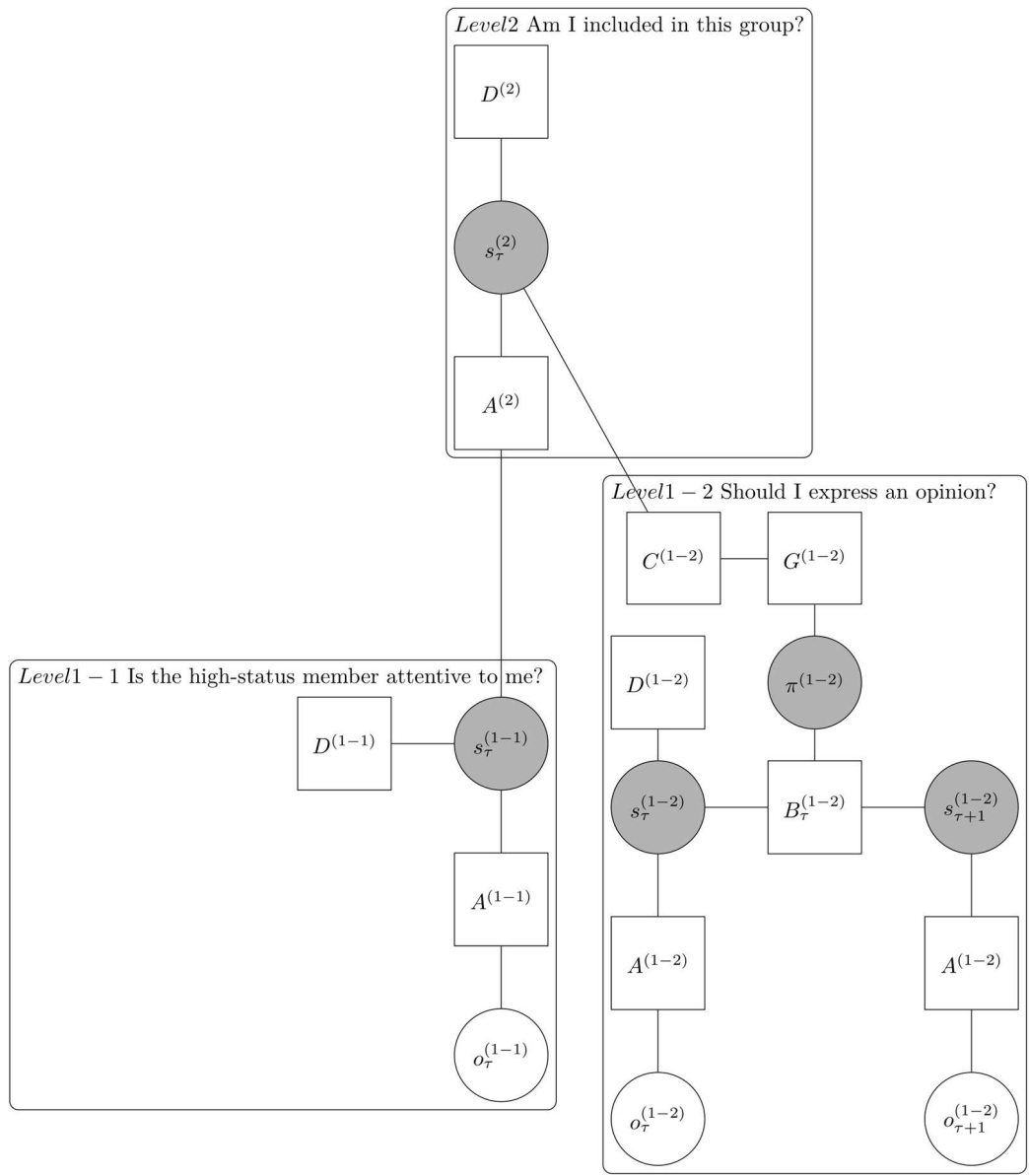

**Fig 3. Bayesian model of the influence of the gaze behavior of a high-status member on the statements of low-status members.**
A: likelihood mapping between states and observations (observation model), B: transition matrices, C: prior preferences, D: initial state priors, G: expected free energy, s: states, o: observations, π: posterior distribution over actions. Superscripts indicate the fields to which the variables belong. The subscript represents the time step. In all fields, the agent of inference is a low-status member.

In Level 1-1, a low-standing member infers whether a high-standing member is attentive based on the gaze behavior of that member. Humans can tell what other people are paying attention to from their gazes [14]. Therefore, an agent receiving a direct gaze infers that the person is attentive to the agent. Alternatively, the agent receiving one's averted gaze infers that the person is not paying attention to the agent. The observational model $p(o^{(1-1)}|s^{(1-1)})$ is described by Eq. (5). $\zeta_{1-1}$ is the precision of Level 1-1 of the observation model. $\sigma()$ is the softmax function. $\exp(-4)$ is added to prevent the formation of $\ln(0)$. By transforming lines 3 and 4 in Eq. (5), the noise associated with the observations and the ambiguities in the mapping between the observations and states can be accounted for.

$$o^{(1-1)} = \{direct\ gaze \quad averted\ gaze\},$$
$$s^{(1-1)} = \{attentive \quad unconcerned\}$$
$$A_{1-1} = \begin{bmatrix} 1 & 0 \\ 0 & 1 \end{bmatrix}$$
$$A^{(1-1)}: p(o^{(1-1)}|s^{(1-1)}) = \sigma(\zeta_{1-1} * \ln(A_{1-1} + \exp(-4))) \qquad (5)$$

The agent infers the mental state of the high-status member at Level 1-1. Therefore, the model evidence of the inference in Level 1-1 corresponds to the sense of the presence of a high-status member.

In Level 2, a low-status member infers whether they are included in a group based on the results of the inference in Level 1-1. An included state corresponds to a high level of psychological safety. Hirak et al. [15] reported that the leader's inclusivity is positively correlated with the psychological safety perceived by other members. A leader's inclusivity is defined as the leader's openness and accessibility in interactions with other members. It is assumed that the agent thinks that they are included in the group from the attention and interests of the group of high-status members. In contrast, it is assumed that the agent thinks that the high-status members' unconcern is exclusive to them. The observational model $p(s^{(1-1)}|s^{(2)})$ is described by Eq. (6).

$$s^{(1-1)} = \{attentive \quad unconcerned\},$$
$$s^{(2)} = \{included \quad excluded\}$$
$$A_2 = \begin{bmatrix} 1 & 0 \\ 0 & 1 \end{bmatrix}$$
$$A^{(2)}: p(s^{(1-1)}|s^{(2)}) = \sigma(\zeta_2 * \ln(A_2 + \exp(-4))) \tag{6}$$

In Level 1-2, a low-status member selects their action: expressing an opinion or not. Expressing an opinion leads to group feedback. It is assumed that the agent learns whether the group agrees with their opinion. Observational model $p(o^{(1-2)}|s^{(1-2)})$ is described as Eq. (7). The observational model varies when the probability of a faulty inference is considered.

$$o^{(1-2)} = \{agreement \quad disagreement\},$$
$$s^{(1-2)} = \{agreement \quad disagreement\}$$
$$A^{(1-2)}: p(o^{(1-2)}|s^{(1-2)}) = \begin{bmatrix} 0.8 & 0.2 \\ 0.2 & 0.8 \end{bmatrix} \tag{7}$$

The agent has prior preferences for future observations associated with action selection at Level 1-2. Prior preferences encode the rewards and aversions gained from the observations. Expressing an opinion involves interpersonal risk. Expressing a dissenting opinion can lead to social exclusion from a group [11]. We established a negative preference for observing *disagreement*. Considering that psychological safety promotes learning behaviors with interpersonal risk [12], we formulated the magnitude of negative preference changes by inclusiveness in the group inferred at Level 2.

It can be modeled that prior preferences depend on context. In this case, the EFE is described by Eq. (8) [16]:

$$G_\pi^{(1-2)} = - \sum_{s^{(1-2)}, o^{(1-2)}} q(o^{(1-2)}, s^{(1-2)}|\pi^{(1-2)}) \ln \frac{q(s^{(1-2)}|o^{(1-2)}, \pi^{(1-2)})}{q(s^{(1-2)}|\pi^{(1-2)})}$$
$$- \sum_{o^{(1-2)}} q(o^{(1-2)}|\pi^{(1-2)}) \ln p(o^{(1-2)}|s_\tau^{(2)}) \tag{8}$$

The second term $p(o^{(1-2)}|s_\tau^{(2)})$ in Eq. (8) is the prior preference, which depends on the context at time τ. When the agent assumes that they are included in the group and that psychological safety is high, the agent perceives it as safe to take interpersonal risks. In contrast, when the agent assumes that they are excluded from the group and psychological safety is low, the agent perceives the danger of taking interpersonal risks.

Therefore, a negative preference is set only for $o^{(1-2)} = disagreement$ when $s^{(2)} = excluded$. Prior preference is described as context-dependent, as in Eq. (9).

$$o^{(1-2)} = \{agreement \quad disagreement\},$$
$$s^{(2)} = \{included \quad excluded\}$$

$$p(o^{(1-2)}|s^{(2)}) = \begin{bmatrix} 0 & 0 \\ 0 & -1 \end{bmatrix} \tag{9}$$

The EFE of Eq. (8) allows for the consideration of prior preferences as a function of the discrete context. In modeling the influence of the level of psychological safety on action selection, we formulated the prior preference as shown in Eq. (10).

$$C^{(2)}: p(o^{(1-2)}|C) = p(s^{(2)} = included)p(o^{(1-2)}|s^{(2)} = included)$$
$$+ p(s^{(2)} = excluded)p(o^{(1-2)}|s^{(2)} = excluded) \tag{10}$$

Then, EFE is described in Eq. (11).

$$G_\pi^{(1-2)} = - \sum_{s^{(1-2)}, o^{(1-2)}} q(o^{(1-2)}, s^{(1-2)}|\pi^{(1-2)}) \ln \frac{q(s^{(1-2)}|o^{(1-2)}, \pi^{(1-2)})}{q(s^{(1-2)}|\pi^{(1-2)})}$$
$$- \sum_{o^{(1-2)}} q(o^{(1-2)}|\pi^{(1-2)}) \ln p(o^{(1-2)}|C) \tag{11}$$

**Simulations based on the Bayesian model**

We performed simulations using the proposed model of the influence of a high-status member's gaze behavior on the low-status members' statements to design the intervention behavior of an avatar robot.

The action selection probability of a low-status member in each condition was simulated. The action selection probability is calculated using Eq. (12).

$$\pi = \{express\ an\ opinion \quad do\ not\ expree\ an\ opinion\}$$
$$p(\pi) = \sigma(-\gamma G) \tag{12}$$

The term $\gamma$ is the expected precision of EFE. The initial value was set to 1. Term G is a vector consisting of the EFEs of the two actions. The term $\sigma()$ is the softmax function. We used the MATLAB code spm_MDP_VB_X_tutorial.m from [7] to perform the simulations.

The avatar robot can intervene in observation in Level 1-1 ($o^{(1-1)}$), likelihood mapping between states and observations in Level 1-1 ($p(o^{(1-1)}|s^{(1-1)})$), and prior in Level 1-1 ($p(s^{(1-1)})$), which express predictions before observation. Conditions that changed the model variables were simulated. The simulation conditions are listed in Table

1. By observing the action selection that follows the EFE, we set the initial value of the EFE's expected precision ($\gamma$) to one.

**Table 1. Conditions of the simulation.**

| Condition | Observation $o^{(1-1)}$ | Likelihood mapping $A^{(1-1)}: p(o^{(1-1)}\|s^{(1-1)})$ | Initial state prior $D^{(1-1)}: p(s^{(1-1)})$ |
|---|---|---|---|
| 1 | Direct gaze | $\zeta_{1-1} = 0.2$ $p(o^{(1-1)}\|s^{(1-1)}) = \begin{bmatrix} 0.691 & 0.309 \\ 0.309 & 0.691 \end{bmatrix}$ | $p(s^{(1-1)}) = \begin{bmatrix} 0.5 \\ 0.5 \end{bmatrix}$ |
| 2 | Averted gaze | $\zeta_{1-1} = 0.2$ $p(o^{(1-1)}\|s^{(1-1)}) = \begin{bmatrix} 0.691 & 0.309 \\ 0.309 & 0.691 \end{bmatrix}$ | $p(s^{(1-1)}) = \begin{bmatrix} 0.5 \\ 0.5 \end{bmatrix}$ |
| 3 | Direct gaze | $\zeta_{1-1} = 0 \sim 1.0$ | $p(s^{(1-1)}) = \begin{bmatrix} 0.5 \\ 0.5 \end{bmatrix}$ |
| 4 | Direct gaze | $\zeta_{1-1} = 0.2$ $p(o^{(1-1)}\|s^{(1-1)}) = \begin{bmatrix} 0.691 & 0.309 \\ 0.309 & 0.691 \end{bmatrix}$ | $a = 0 \sim 1.0$ $p(s^{(1-1)}) = \begin{bmatrix} a \\ 1-a \end{bmatrix}$ |

s: states, o: observations, $\zeta_{1-1}$: precision of Level 1-1 the likelihood mapping. Superscripts indicate the fields to which the variables belong.

In condition 1, we simulated the action selection probability when the agent observes the direct gaze of the high-status member as observation $o^{(1-1)}$. We set the precision of the observation model to 0.2 to account for the uncertainties in the mapping between the inferred state and observed gaze. We set the prior as $p(s^{(1-1)}) = \begin{bmatrix} 0.5 \\ 0.5 \end{bmatrix}$ under the assumption that the agent equally predicts whether the high-standing member pays attention to and is interested in them.

In condition 2, the action selection probability was simulated when the agent observes the averted gaze of a high-status member as observation $o^{(1-1)}$. By comparing the results of conditions 1 and 2, the influence of the avatar robot's conversion of an averted gaze to a direct gaze on a low-status member's statement was predicted. We set the precision of the observation model prior to the same values as condition 1.

Under condition 3, the action selection probability was simulated by changing the precision of the observation model $p(o^{(1-1)}|s^{(1-1)})$. The prior was set as $p(s^{(1-1)}) = \begin{bmatrix} 0.5 \\ 0.5 \end{bmatrix}$ under the assumption that the agent equally predicts whether the high-status member pays attention to and is interested in them. The influence of a high-status member changing their gaze behavior to one that expressed more attention and interest in what the low-status member said was predicted.

In condition 4, the action selection probability was simulated by changing the prior $p(s^{(1-1)})$. The precision of the observational model was set to 0.2 to account for uncertainties in the mapping between the inferred state and observed gaze. The prior encodes the prediction of whether the high-status member pays attention and interest to the agent. The influence of adding a prior motion [17] to the gaze behavior of the high-status member by the avatar robot, which enabled the prediction of the gaze movement on what the low-status member said, was predicted.

Table 2 lists the simulation results of action selection probability under conditions 1 and 2. The probability of expressing an opinion is higher under condition 1 than under condition 2. Thus, it is predicted that changing the gaze behavior of the high-status member from an averted gaze to a direct gaze will encourage the low-status member to speak.

Fig 4 shows the simulation results of the action selection probability under condition 3. The higher the precision of the observation model, the higher the probability of expressing an opinion. Therefore, it is predicted that changing the gaze behavior of high-status members to one that expresses more attention and interest will encourage a low-status member to express an opinion.

Fig 5 shows the simulation results of the action selection probability under condition 4. The higher the degree of prediction that a high-status member will pay attention and show interest in the agent, the higher the probability of an opinion being expressed. Based on this result, it is predicted that adding preliminary movements to the gaze behavior of high-status members would encourage low-status members to speak.

Based on these findings, we designed the gaze behavior of an avatar robot. In this study, we designed the gaze behavior when the remote participant spoke, which seemed to express more attention to local participants.

**Table 2. Simulation results of action selection probability under conditions 1 and 2.**

s: states, o: observations, $\zeta_{1-1}$: precision of Level 1-1 the likelihood mapping.

| Condition | Observation $o^{(1-1)}$ | Likelihood mapping $A^{(1-1)}: p(o^{(1-1)}\|s^{(1-1)})$ | Initial state prior $D^{(1-1)}: p(s^{(1-1)})$ | Probability of expressing an opinion |
|---|---|---|---|---|
| 1 | Direct gaze | $\zeta_{1-1} = 0.2$<br>$p(o^{(1-1)}\|s^{(1-1)}) = \begin{bmatrix} 0.691 & 0.309 \\ 0.309 & 0.691 \end{bmatrix}$ | $p(s^{(1-1)}) = \begin{bmatrix} 0.5 \\ 0.5 \end{bmatrix}$ | 0.3729 |
| 2 | Averted gaze | $\zeta_{1-1} = 0.2$<br>$p(o^{(1-1)}\|s^{(1-1)}) = \begin{bmatrix} 0.691 & 0.309 \\ 0.309 & 0.691 \end{bmatrix}$ | $p(s^{(1-1)}) = \begin{bmatrix} 0.5 \\ 0.5 \end{bmatrix}$ | 0.0571 |

Superscripts indicate the fields to which the variables belong. Using a mathematical model, we simulated the probability of a low-status member speaking after seeing the gaze behavior of a high-status member in each condition.

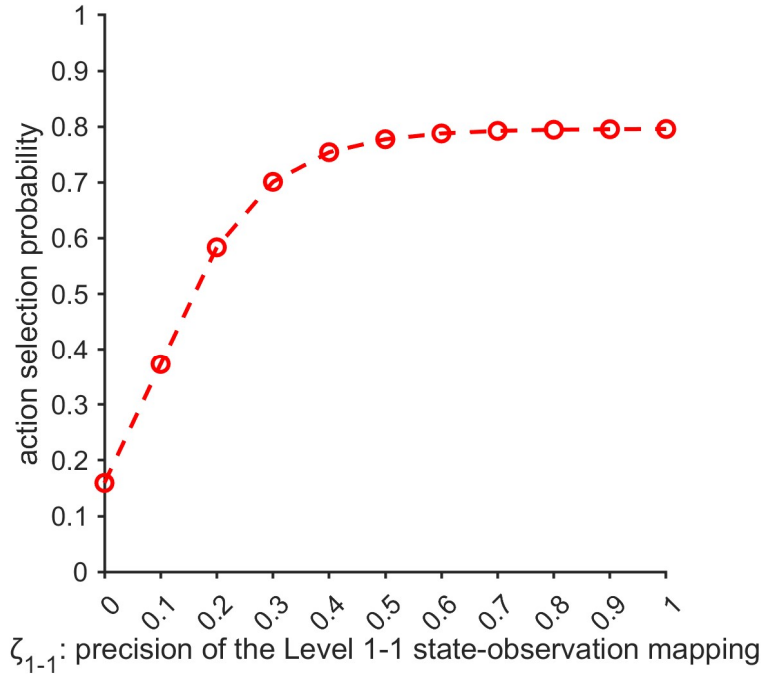

**Fig 4. The influence of precision of the Level 1-1 observation model on the probability that a low-status member expresses an opinion.**

Precision adjusts the ambiguity of the mapping between states and observations. Higher precision indicates that the mapping is more precise. The observation model in Level 1-1 is a mapping between the attentional states of high-status members (attentive/unconcerned to a low-status member) and their gaze behavior.

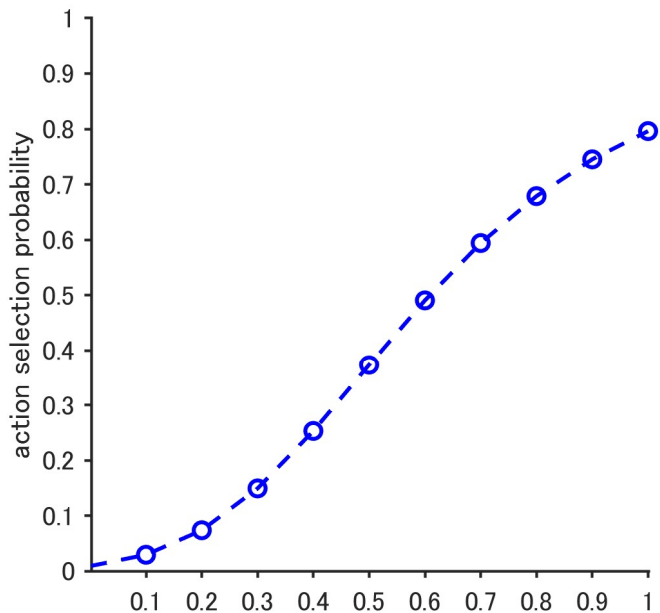

**Fig 5. The influence of prior distribution in Level 1-1 $(p(s^{(1-1)}) = [a \quad 1-a])$ on the probability that a low-status member expresses an opinion.**

The state derived by a low-status member in Level 1-1 is the attention state of a high-status member (attentive/unconcerned to a low-status member). A higher value of *a* means that a low-status member predicts that a high-status member is more likely to direct their attention and interest toward a low-status member.

### Design of the gaze behavior of the avatar robot

Based on the simulation results, the gaze behavior of an avatar robot was designed in conjunction with a remote participant's speech. First, the requirements for gaze behavior are explained. Second, the design of a gaze behavior is explained that satisfies these requirements. Then, the actual designed gaze behavior is explained.

Based on the initial simulation results, we determined to make the avatar robot automatically cast a direct gaze, regardless of the remote participant's gaze behavior. However, if the gaze behavior appears to be automated, the local participants may not perceive that the remote participant is paying attention to or is interested in them. Thus, the first requirement is that gaze behavior should be natural and not artificial. We also predicted that gaze behavior expressing more attention, interest, and preliminary movement would encourage low-status members to speak in the simulations. Therefore, the second requirement is that gaze behavior should express more attention and interest. We used a preliminary movement to satisfy this requirement.

Next, the design of a gaze behavior is explained that satisfies these two requirements. In this study, natural behavior is designed by using the techniques and ideas of animators. Animators express vivid behaviors by emphasizing and exaggerating necessary behavioral features and omitting unnecessary behavioral features [17]. Therefore, our design guidelines emphasize behavioral features that can be a cue to pay attention for local participants and omit the rest. It is likely that this guideline realizes a gaze behavior that satisfies both requirements.

Finally, the designed gaze behavior is described. Table 3 lists the behavioral features incorporated into the gaze behavior of the avatar robot and their meanings. We implemented these behavioral features in the gaze behavior while a remote participant was speaking.

**Table 3. Behavioral features incorporated into the gaze behavior of the avatar robot and meanings of them.**

| Behavioral features incorporated into the gaze behavior of the avatar robot | Meanings of the behavioral features |
|---|---|
| Blinks before and after the head swing | ● Preliminary movement of viewpoint swift [17][18].<br>● Imitation of real behavior [17].<br>● Emphasis on head swing stoppage [17]. |
| Omitting the blink except for the blink before and after the head swing | ● Enhancing impressions of blinks before and after the head swing.<br>● Enhancing impressions of direct gaze [19]. |
| Emphasis on slow-in/slow-out [17] in head swing | ● Imitation of real behavior.<br>● Emphasis on head swing stop [17]. |
| Implementation of eye movements associated with blinking and head swinging | ● Imitation of real behavior. |
| Head swinging associated with a breather | ● Imitation of real behavior.<br>● Expression that the subjects of the speech and eye behavior are the same |

# Experiment: Effect of the intervention in the gaze behavior of the avatar robot on the statements of local participants

## Gaze behavior of the avatar robot used in the experiment

Participants were divided into an experimental and a control group. For each group, we used the different gaze behavior of the avatar robot when the remote participant (the researcher) spoke.

For the experimental groups, the avatar robot described in the previous section was used. In this task, the researcher acted as a remote participant using an avatar robot, and the two participants were local participants. Fig 6 shows the arrangement of the participants and the avatar robot in the experiment. The avatar robot looked at each participant in the experimental group in turn. When the researcher breathed, the avatar robot changed the direction of its gaze.

In the control group, the avatar robot looked at the center of the two participants throughout the researcher's speech. To avoid an unnatural impression, we implemented and used a blink pattern in which the avatar robot blinked once after three seconds and twice after another three seconds.

The specific implementation of the gaze behaviors is described in the Supporting Information.

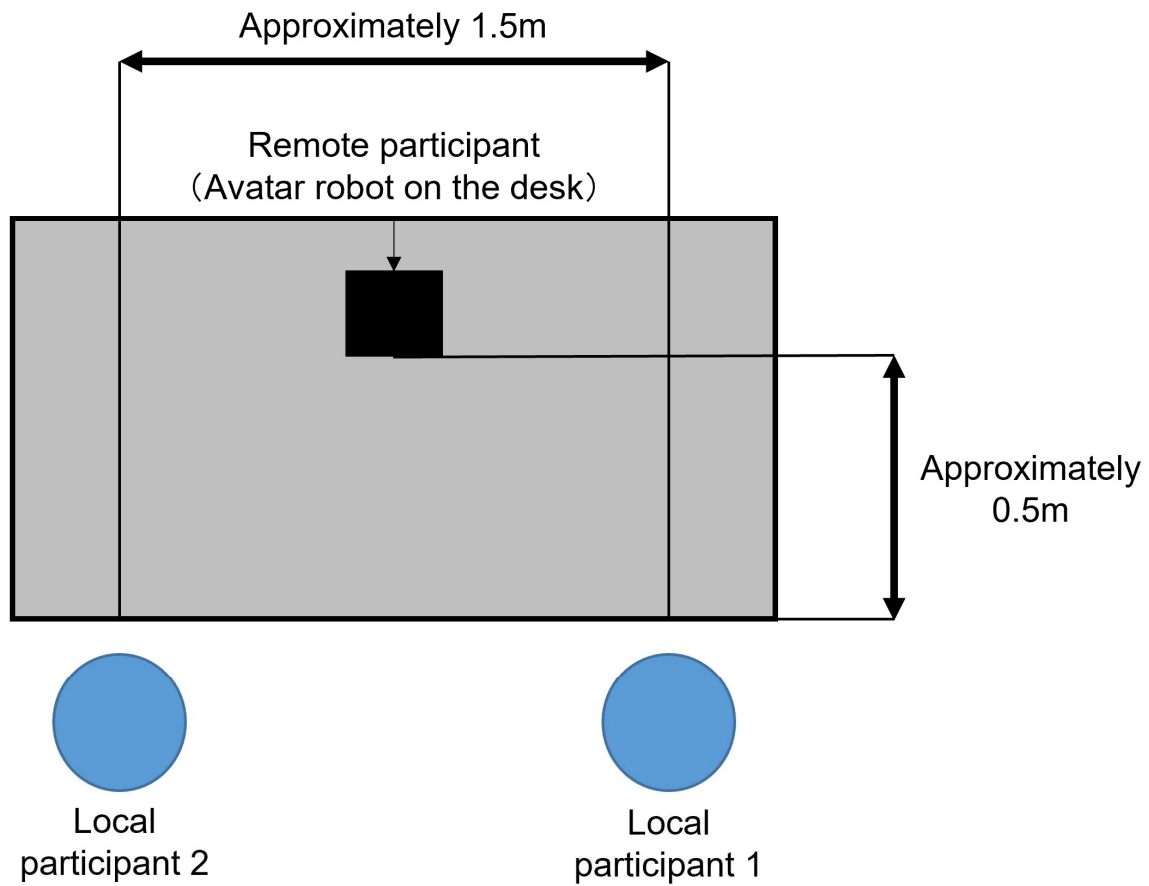

**Fig 6. Arrangement of participants and the avatar robot in the experiment.**

## Participants

Participants were recruited from among students at the University of Tokyo who had no visual or hearing impairments and whose native language was Japanese, using both snowball sampling and voluntary responses. Twenty-two students (16 male, six female) in the experimental group and 22 students (16 male, five female, one no answer) participated in this study. For controlling the influence of acquaintances and gender combinations, we created pairs of same-gender participants who did not know each other. A female pair who was acquainted with each other and a pair of female and no-answer participants who had selected the same candidate before the discussion were excluded from the analysis.

This study was approved by the Research Ethics Committee of the School of Engineering, The University of Tokyo (Approval No. KE22-59). All study participants provided informed consent.

## Design

We used a hidden-profile task [8] to quantitatively investigate whether local participants' statements were promoted. In the hidden-profile task, participants received partial information about their choices. Participants can only choose the best option if they share unique information with each other. Therefore, a hidden-profile task is used to study group information elaboration [13] and team performance in learning [20].

A group consisting of one researcher and two participants performed a hidden profile task. Participants read the profiles of three hypothetical candidates for the student body president and then decided which candidate was best suited for the position. The profiles did not include complete information about the candidates and contained unique information provided to only one participant. We examined whether each participant's statement was promoted by evaluating the amount of unique information shared.

Participants played a game that reproduced the relationship between their boss and subordinates during discussions in the hidden-profile task. We call this the boss–subordinate game. The researcher played the game in the role of the boss, and participants acted as subordinates. We calculated each participant's score using the boss-subordinate game rules. Participants were then told that in addition to the gratuities, they would receive snacks according to their scores. After the experiment, all participants were given the same number of snacks. The boss-subordinate game states that a subordinate who disagrees with their boss has the risk of a point deduction for their statements. This risk is related to the subjective judgment of the boss. This rule structure reproduces a situation in which low-status members are reluctant to express their opinions because of the interpersonal risks.

The profiles were designed so that each participant supported a different candidate: candidate B or candidate C. The researcher supported candidate B as the supervisor. Thus, a participant who supported candidate C risked their statements according to the boss-subordinate game rules. The profiles were created by modifying the method used in [8].

In the actual experiments, there were cases where the participants did not support the candidates as intended. In such cases, the researcher selected and supported candidates B or C, so there were participants who agreed and disagreed with the boss. We refer to a participant who agrees with the boss as Participant 1 and a participant who disagrees with the boss as Participant 2.

## Procedure

After completing the informed consent form, the participants completed questionnaire testing personality traits prior to the experiment. They then began the hidden-profile task. This hidden-profile task consists of individual and discussion phases. In the individual phase, each participant reads the profile and indicates which candidates they support at that time.

Then, the researcher explains the boss-subordinate game that is played in the discussion phase. The researcher then moved to another room and connected to an avatar robot. After confirming that the connection to the avatar robot was successful, the researcher asked the participants to report the candidates they had selected for the individual phase. The order of reporting was as follows: participants who received the profile that recommended candidate B reported first, followed by those who received the profile that recommended candidate C.

The researcher then explained the discussion phase and gave an opinion as the boss (which candidate was supported and why). While the researcher explained this, the avatar robot looked at each participant in turn in the experimental group. In the control group, the avatar robot continued to gaze at the middle of the two participants, while the researcher gave an opinion. After the researcher expressed an opinion as the boss, the participants engaged in a free discussion about which candidate should be nominated for student body president. During the free discussion, the researcher did not speak, and the avatar robot continued to observe the center of the two participants.

The participants reported to the researcher the candidate who should be nominated as student body president after reaching an agreement within five minutes of the start of the free discussion. The hidden profile task was completed when the researcher received the report. The researcher distributed a questionnaire asking for unique information shared by the group during the free discussion, as well as subjective ratings of the discussion phase experience. After debriefing and conducting the interviews, the participants received snacks as an additional reward and were dismissed.

## Measures

Objective and subjective indexes were used to assess the effect of the avatar robot's gaze behavior on participants.

As an objective index, the amount of unique information shared by each participant during the free discussion was measured. The amount of unique information shared by Participant 1 (who agreed with the researcher as the boss) was referred to as sum1. The amount of unique information shared by Participant 2 (who disagreed with the researcher as the boss) was referred to as sum2. The profile that Participant 1 had read contained six unique items of information that supported the boss's opinion and three items of information that did not support the boss's opinion. The amount of shared unique information that supported the boss's opinion was measured, and this number is referred to as agreement1. That unique information indicated that the candidate supported by the boss was desirable or indicated that the candidate not supported by the boss was undesirable. Additionally, the amount of shared unique information that did not support the boss's opinion was measured, and this number was referred to as disapproval1. The total number of agreement1 and disapproval1 responses was sum1. Participant 1 answered questions about the unique information they had shared with the group during the free discussion from all the unique information. Participant 2 read nine unique pieces of information that did not support the boss's opinion. Participant 2 answered questions about the unique information they shared with the group during the free discussion from all the unique information.

In addition, the amount of unique information that the participants were hesitant to share was measured. The number of responses from Participant 1 was called hesitation1, and the number of responses from Participant 2 was called hesitation2.

Subjective indexes to measure the effects on participants were used that were not included in the objective indexes. The subjective indexes were based on a model of the influence of a high-ranking member's gaze behavior on the opinions of low-ranking members.

How easy it was to identify the gaze direction while the researcher was connected to the avatar robot was measured. It was checked whether participants could observe the direct gaze of the high-ranking member (the researcher) at Level 1-1 of the model.

Perceptions of attention and interest, while the researcher was connected to the avatar robot, were measured. Perceptions of attention and interest indicate whether participants infer that they are receiving attention and interest from the high-ranking member in Level 1-1 of the model. Perception of attention and interest is an index of how large the model evidence is for the model from which participants inferred that they received attention and interest from high-ranking members (i.e., how large the presence of high-ranking members is).

We then measured the inclusiveness of each group. We checked whether participants in Level 2 inferred from their perception of attention and interest that they were included in the group.

The ease with which participants were able to express their opinions during the free discussion was measured. It was checked whether participants felt that it was easy to select the action of expressing themselves in Level 1-2.

Participants responded on a 7-level Likert scale to four of the above items: ease of identifying gaze direction, perception of attention and interest, involvement in the group, and ease of expressing opinions.

It was assumed that the results would be influenced by the participants' personality traits. Therefore, prior to the hidden profile task, participant's extraversion and neuroticism using the Big-Five marker was measured [21].

After debriefing, participants were asked about whether they perceived any risk in sharing information during the free discussion and what their impression of the avatar robot was while the researcher was speaking.

In addition, a video of the discussion phase of the hidden profile task was recorded and used to discuss the experimental results.

**Hypothesis**

The gaze behavior used in the experimental group was designed to express attention and interest to local participants. Thus, we hypothesized the following:

H1. The ease with which gaze direction could be identified increased in the experimental group compared to the control group for both Participant 1 (a participant who agreed with the researcher as the boss) and Participant 2 (a participant who disagreed with the researcher as the boss).

H2. The perception of attention and interest increased in the experimental group compared to the control group for both Participants 1 and 2.

Based on the model of the influence of the gaze behavior of the high-status member on the statements of low-status members, we hypothesized the following:

H3. Participant 2's inclusivity in the experimental group increased compared to the control group.

H4. The ease with which participant 2 expresses their opinion increased in the experimental group compared to the control group.

There seemed to be no significant differences between the two groups in terms of Participant 1's inclusion in the group and the ease of expressing their opinions. This is because Participant 1 agreed with the boss and rated the boss highly in both groups.

In addition, we hypothesize the following:

H5. Index sum2 will increase in the experimental group compared to the control group.

H6. Index hesitation 2 will decrease in the experimental group compared to the control group.

This was due to the fact that the designed gaze behavior promoted the statements of Participant 2's statements with interpersonal risks. In contrast, Participant 1 had low risk in their statements because they contained a lot of clear information that supported the boss's opinion. There were no significant differences between the two groups in terms of sum1 and hesitation1.

**Data analysis**

We analyzed the results of the experiments with 40 participants (experimental group: 16 male and four female; control group: 16 male and four female). A between-subjects analysis was performed. When the personality traits of the participants (extraversion and neuroticism) were significantly related to the indices, we performed an analysis of covariance (ANCOVA) with personality traits as covariates and the participant group of (experimental group or control group) as a factor. When the effect of the personality traits of the participants on the indices was not significant, a two-tailed t-test was performed to compare the experimental and control groups on the objective and subjective ratings. The threshold for significance (alpha) was 0.05.

# Results

Fig 7 shows the amount of unique information shared by a participant who agreed with the researcher as boss (Participant 1). Fig 8 shows the amount of unique information shared by a participant who disagreed with the researcher as boss (Participant 2). The index sum2 was significantly lower in the experimental group than in the control group ($t(18) = -3.104$, $p = 0.006$). Index sum2 shows the amount of unique information that Participant 2 shared. This result contradicts H5. Moreover, the value of sum1 was also significantly lower in the experimental group than in the control group ($t(18) = -2.137$, $p = 0.047$). The index sum1 shows the amount of unique information shared by Participant 1.

There were no significant differences between the experimental and control groups for hesitation 2 ($t(18) = 0.911$, $p = 0.828$). Thus, H6 was not supported.

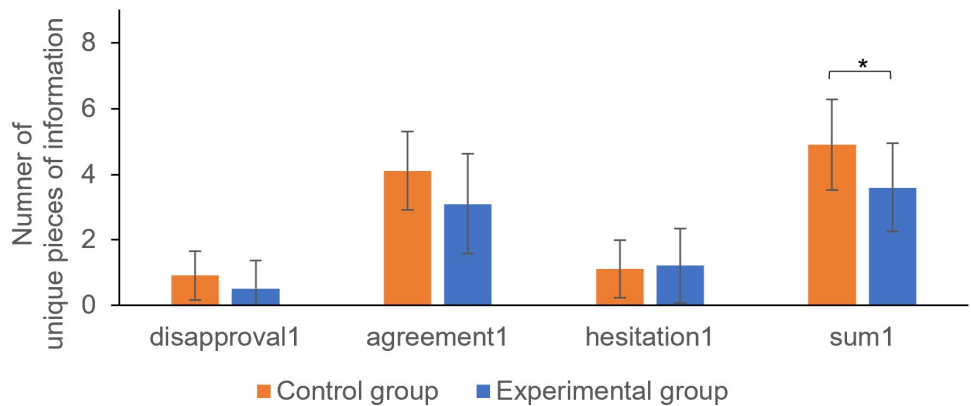

**Fig 7. Number of unique pieces of information shared by a participant who agreed with the high-standing member.** * $p<0.05$, ** $p<0.01$. Error bars indicate standard deviations.

Disapproval1: number of unique pieces of information shared that did not support the opinions of high-status members. Agreement1: number of unique pieces of information shared that supported high status member's opinion. Hesitation1: number of unique pieces of information that were not shared when participants hesitated whether or not to share. Sum1: total of dessent1 and assent1. A participant who agreed with a high status member had six unique pieces of information that supported the high-status member's opinion and three that did not support the high-status member's opinion.

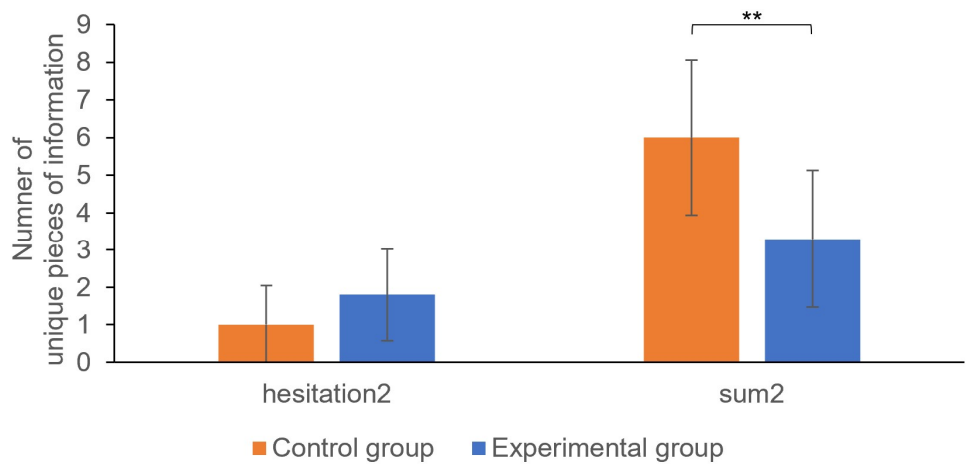

**Fig 8. Number of unique pieces of information shared by a participant who disagreed with the high-status member**. * $p<0.05$, ** $p<0.01$. Error bars show standard deviations.

Hesitation2: number of unique pieces of information that were not shared when participants hesitated whether or not to share. Sum2: number of unique pieces of information shared by a participant who disagreed with the high-status member. A participant who disagreed with the high-status member had nine unique pieces of information that did not support the high-status member's opinion.

Fig 9 shows the subjective rating of Participant 1. Fig 10 shows the subjective rating of Participant 2. For both Participant 1 and Participant 2, the ease of identifying the gaze direction was significantly higher in the experimental group than in the control group (ease of identifying the gaze direction of Participant 1: $t(18) = 5.513$, $p < 0.001$, ease of identifying the gaze direction of Participant 2: $t(18) = 2.994$, $p = 0.008$). These results are consistent with H1.

For both Participant 1 and Participant 2, perceptions of attention and interest were significantly higher in the experimental group than in the control group (Participant 1's perception of attention and interest: $t(18) = 2.832$, $p = 0.011$. Participant 2's perception of attention and interest: $t(18) = 2.374$, $p = 0.029$). Thus, H2 is supported.

However, there were no significant differences between the experimental and control groups in terms of in-group inclusivity or ease of expressing opinions (inclusivity for Participant 1: $t(13.072) = -0.397$, $p = 0.698$; ease of expressing opinions for Participant 1: $t(18) = -1.698$, $p = 0.107$; inclusivity for Participant 2: $t(18) = -0.616$, $p = 0.546$; ease of expressing opinions for Participant 2: $t(18) = -0.843$, $p = 0.410$). Therefore, H3 and H4 are not supported.

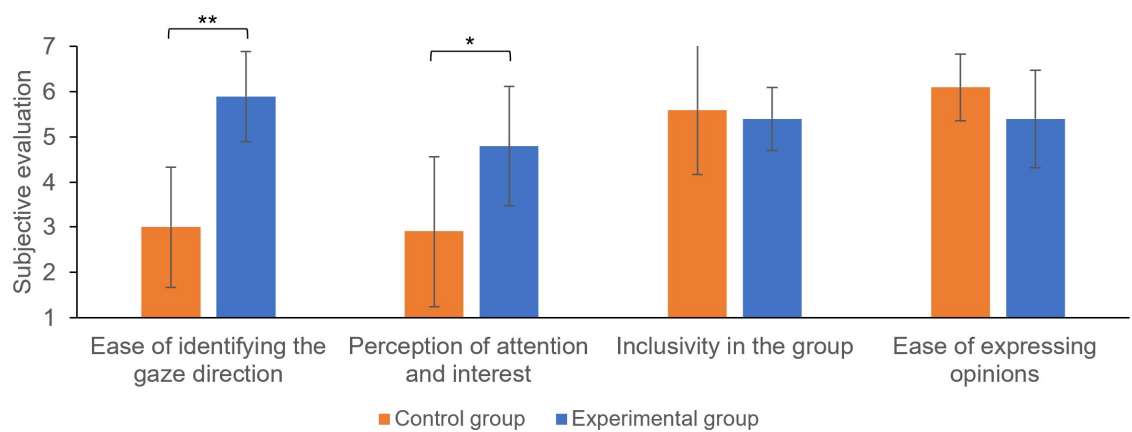

**Fig 9. Subjective evaluation of a participant agreeing with the high-status member.** * $p<0.05$, ** $p<0.01$. Error bars show standard deviations.

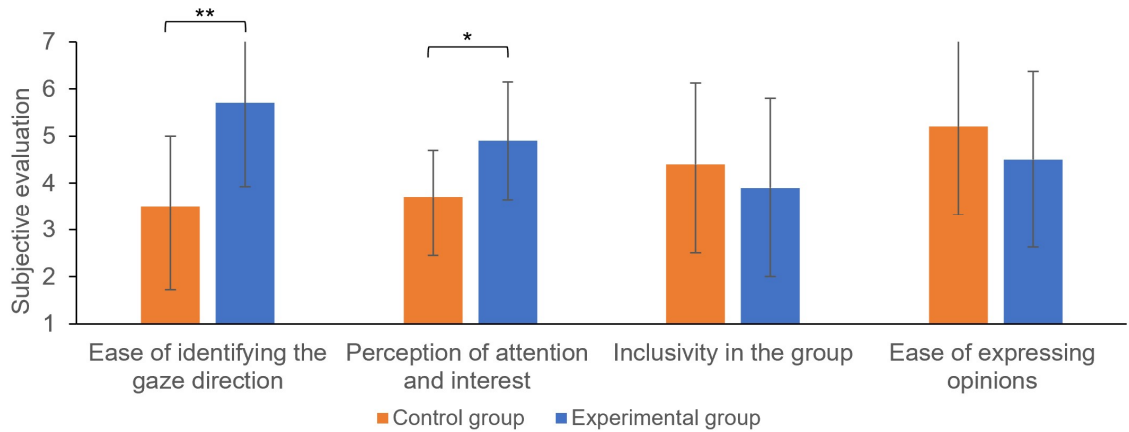

**Fig 10. Subjective evaluation of a participant disagreeing with the high-status member.** * *p*<0.05, ** *p*<0.01. Error bars show standard deviations.

# Discussion

For both Participant 1 and Participant 2, perceptions of attention and interest were significantly greater in the experimental group than in the control group (Participant 1: agrees with the researcher as the boss, Participant 2: disagrees with the researcher as the boss). Therefore, we assumed that the gaze behavior of the avatar robot increased the presence of the remote participant (the researcher as boss), as hypothesized. We also received responses from participants such as "I felt like I was being watched," and "I assumed the researcher was talking to the participant he was looking at" during the post debriefing interview. We hypothesize that the gaze behavior increased the model evidence of each local participant, by inferring that the boss was paying attention and interest to them, and thus increasing the presence of the remote boss.

However, sum1 and sum2 are significantly smaller in the experimental group than in the control group, contrary to H5 (Index sum1 and sum2 show the number of unique pieces of information shared by Participants 1 and 2). We believe that these results are attributable to the incorrect modeling of Level2 in the constructed model.

In the subordinate game, Participant 2 was at risk of a point reduction if they expressed an opinion. If the boss chose the "Rejection of an opinion" option when Participant 2 expressed a dissenting opinion, Participant 2's score would be reduced. Therefore, the presence of a boss may have led Participant 2 to be aware that they were being monitored by the boss and to strongly perceive the risks of expressing an opinion. In other words, we assume that Participant 2 inferred from the boss's attention and interest at Level 2 of the model that the boss was monitoring them.

Based on this discussion, the model of the influence of a high-status member's gaze behavior on low-status members' statements was updated, as shown in Fig 11.

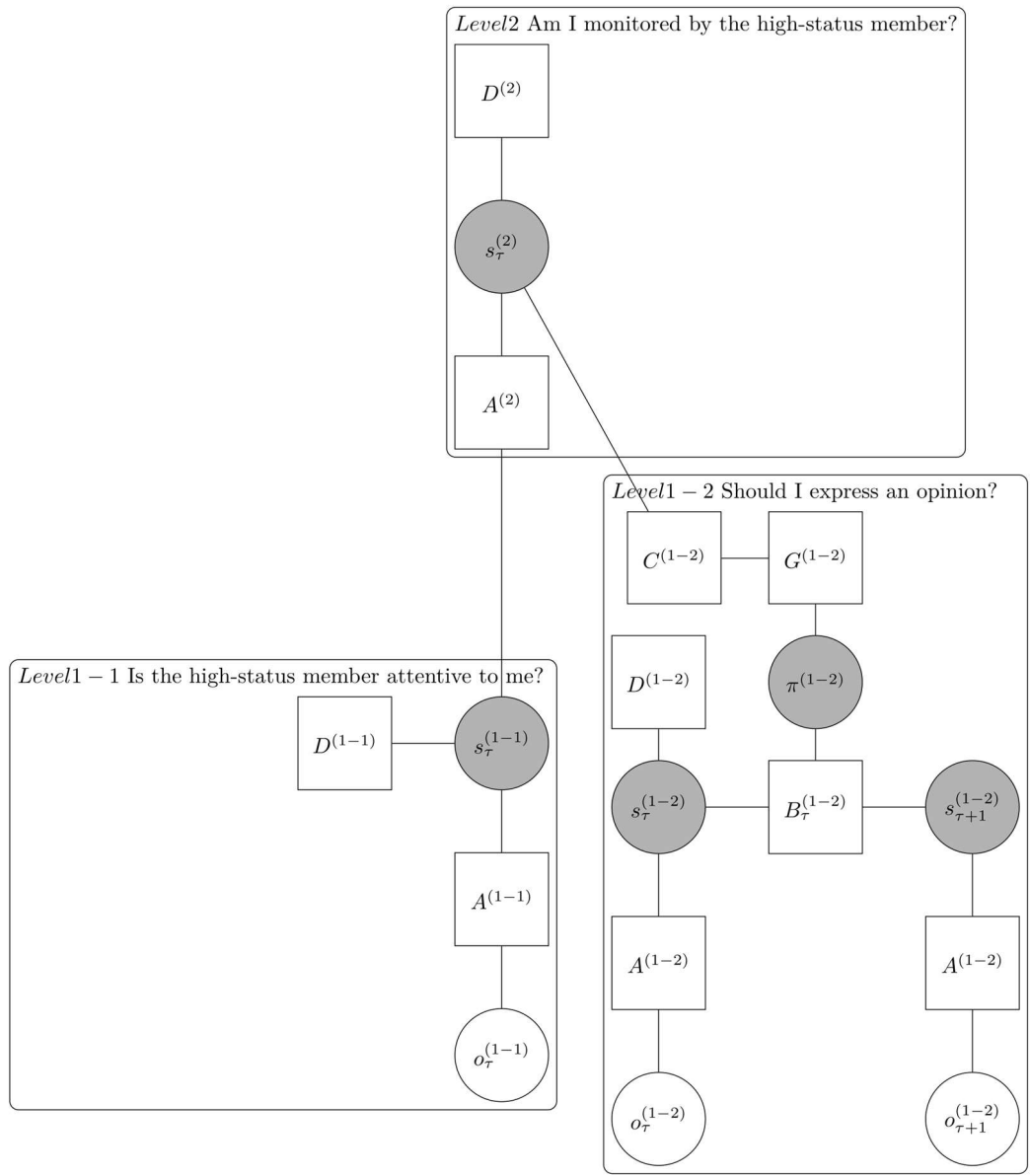

**Fig 11. Modified mathematical model of the influence of the gaze behavior of a high-status member on the statements of low-status members.** Inferred states and observation model in Level 2 is modified. A: likelihood mapping between states and

observations (observation model), B: transition matrices, C: prior preferences, D: initial state priors, G: expected free energy, s: states, o: observations, π: posterior distribution over actions. Superscripts are used to indicate the fields to which variables belong. Subscripts indicate the time step. In all fields, the agent of inference is a low-status member.

First, we explain Level 1-1. In this experiment, perceptions of attention and interest were increased by the gaze behavior of the avatar robot, as hypothesized. Therefore, the observation model for Level 1-1 is defined in Eq. (13), which is the same as before the modification.

$$o^{(1-1)} = \{direct\ gaze \quad averted\ gaze\},$$
$$s^{(1-1)} = \{attentive \quad unconcerned\}$$

$$A_{1-1} = \begin{bmatrix} 1 & 0 \\ 0 & 1 \end{bmatrix}$$
$$p(o^{(1-1)}|s^{(1-1)}) = \sigma(\zeta_{1-1} * \ln(A_{1-1} + \exp(-4))) \tag{13}$$

Next, Level 2 is described. There was no significant difference between the experimental and control groups in terms of in-group inclusivity. However, we believe that the perception of the boss's attention and interest increased the feeling of being monitored by the boss. Therefore, we changed the hidden state $s^{(2)}$ from $s^{(2)} = \{included \quad excluded\}$ to $s^{(2)} = \{monitored \quad not\ monitored\}$. We reformulate the observation model in Level2 as Eq. (14).

$$s^{(1-1)} = \{attentive \quad unconcerned\},$$
$$s^{(2)} = \{monitored \quad not\ monitored\}$$

$$A_2 = \begin{bmatrix} 1 & 0 \\ 0 & 1 \end{bmatrix}$$
$$p(s^{(1-1)}|s^{(2)}) = \sigma(\zeta_2 * \ln(A_2 + \exp(-4))) \tag{14}$$

Next Level 1-2 is described. The observation model for Level 1-2 is defined in Eq. (15), which is the same as before the modification.

$$o^{(1-2)} = \{agreement \quad disagreement\},$$
$$s^{(1-2)} = \{agreement \quad disagreement\}$$

$$p(o^{(1-2)}|s^{(1-2)}) = \begin{bmatrix} 0.8 & 0.2 \\ 0.2 & 0.8 \end{bmatrix} \tag{15}$$

We changed the prior preference in Level 1-2 because the context $s^{(2)}$ is modified. In the experimental task, being monitored by the boss led low-status members (participants) to perceive the risks of a point deduction according to the boss-subordinate game rule. Thus, the boss-subordinate game rule determines the risk of disagreeing with the boss. Therefore, we changed $s^{(2)}$ from $s^{(2)} = \{included \quad excluded\}$ to $s^{(2)} = \{monitored \quad not\ monitored\}$ and set a negative preference only to $o^{(1-2)} = disagreement$ when $s^{(2)} = monitored$. We reformulate the prior preference at Level 1-2 as Eq. (16).

$$o^{(1-2)} = \{agreement \quad disagreement\},$$
$$s^{(2)} = \{monitored \quad not\ monitored\}$$

$$p(o^{(1-2)}|s^{(2)}) = \begin{bmatrix} 0 & 0 \\ -1 & 0 \end{bmatrix} \tag{16}$$

The simulations were performed with the updated model. We changed observation $o^{(1-1)}$ under each condition. Table 9 lists the simulation conditions and results of the action selection probability. The model variables that did not change in each condition were set the same as those in the model before modification and are shown in the Supporting Information.

**Table 9. Conditions of the simulation using the modified model and simulation results of action selection probability under each condition.**

| Condition | Observation $o^{(1-1)}$ | Likelihood mapping $A^{(1-1)}: p(o^{(1-1)}|s^{(1-1)})$ | Initial state prior $D^{(1-1)}: p(s^{(1-1)})$ | Probability of expressing an opinion |
|---|---|---|---|---|
| 1 | Direct gaze | $\zeta_{1-1} = 0.2$<br>$p(o^{(1-1)}|s^{(1-1)})$<br>$= \begin{bmatrix} 0.691 & 0.309 \\ 0.309 & 0.691 \end{bmatrix}$ | $p(s^{(1-1)}) = \begin{bmatrix} 0.5 \\ 0.5 \end{bmatrix}$ | 0.0571 |
| 2 | Averted gaze | $\zeta_{1-1} = 0.2$<br>$p(o^{(1-1)}|s^{(1-1)})$<br>$= \begin{bmatrix} 0.691 & 0.309 \\ 0.309 & 0.691 \end{bmatrix}$ | $p(s^{(1-1)}) = \begin{bmatrix} 0.5 \\ 0.5 \end{bmatrix}$ | 0.3729 |

s: states, o: observations, $\zeta_{1-1}$: precision of Level 1-1 the likelihood mapping. Superscripts indicate the fields to which the variables belong.

The simulation results in Table 9 indicate that the probability of speaking up in condition 1 is lower than that in condition 2, so the high-status member (or the avatar robot) giving a direct gaze to the low-status members withholds their statements. The experimental result that sum2 was lower in the experimental groups than that in the control groups was supported by simulations with the updated model.

We believe that Participant 2's withholding of statements led to the reduction of sum1 in the experimental groups. When Participant 2 did not say very much, Participant 1 did not need to state much about the object. In addition, a significant positive correlation was found between sum2 and sum1 ($r = 0.587$, $N = 20$, $p = 0.007$).

The gaze behavior of the avatar robot was only performed before the free discussion, in which participants expressed their opinions. Nevertheless, gaze behavior significantly increased the presence of remote participants and influenced the statements of local participants in the free discussion.

## Conclusion

In this study, we formulated sense of presence as model evidence for inferring the mental states of others. We then proposed that gaze behavior should increase the sense of presence, which is the model evidence, to infer that the remote participant is attentive to the local participants. Simulations following the model predicted that gaze behavior would promote the local participants' statements. We investigated whether the designed gaze behavior promoted local participants' expressing an opinion using the meeting task that reproduced the discussion of a group with different standings. The experimental results suggest that the gaze behavior of the avatar robot significantly increases the presence of the remote participant, which restrains local participants' from expressing an opinion in the context of the experimental task.

The gaze behavior of the proposed avatar robot increased the presence of remote participants. We believe that gaze behavior can contribute to communication in a context where the presence of the remote participants has a positive effect. For example, when subordinates want to appeal to their skills or make a contribution to their boss, the presence of the boss will encourage the subordinates' statements. In these cases, transforming the boss's behavior into the avatar robot's gaze behavior developed in this study may promote the subordinates' statements.

The finding that less attentive behavior while speaking promotes subordinates' statements in this context can also be used for intervention by an avatar robot. The behavior of an avatar robot can change in this context. The avatar robot may exhibit less attentive behavior in contexts where the presence of a boss exerts pressure and may exhibit the gaze behavior developed in this study in contexts where the presence of a boss encourages subordinates. Therefore, it is important to understand the appropriate presence of remote participants in each context.

The experimental results showed that the presence of the remote participants restrained the local participants' statements. This result is inconsistent with our hypothesis. However, this indicates that the presence of the model has sufficient influence to constrain the participants' statements. In addition, previous studies [1][2] have taken the approach of increasing presence. However, as the current study shows, presence can have a negative effect on communication. Therefore, we argue that it is worthwhile to study the influence of presence and control methods using mathematical models.

This study has some limitations. First, the experiment was conducted with only Japanese participants. The effects of an avatar robot's gaze behavior may be different in different cultures. Second, participants' prior knowledge of the robot's features was not standardized in this experiment. It is hypothesized that the perception of behavior will be different when participants believe that the robot reflects behavior of the remote participants and when they do not.
.

## Acknowledgments

We would like to express our sincere gratitude to Professor Tamotsu Murakami of the Department of Mechanical Engineering, School of Engineering, University of Tokyo, for his valuable comments during the workshop.

We express our sincere gratitude to Ryuichi Suzuki, Mari Yasuda, Naoki Nishida, Shin Shiroma, and other Sony Group employees for providing us with the necessary equipment for our research activities, consulting with the code used to implement the avatar robot's behavior, and for the helpful suggestions through meetings and the avatar robot experience.

We would like to express our sincere gratitude to Kazuhiro Oikawa, Principal Technical Specialist, for advising us on the facilities, equipment, and experimental environment and for preparing the research environment.

We would like to express our sincere thanks to Secretary Yuri Nozaki for her help in handling the paperwork for the experimental participants' rewards.

We would also like to express our sincere appreciation to all those who participated in the preliminary and main experiments of this study as well as those who helped us recruit participants.

# Supporting Information

**S1 Text. Specific Implementation of the gaze behavior of the avatar robot.**

Data was collected on the changes in head rotation, gaze, and blinking that accompany gaze behavior. The data were analyzed according to the situation. S2 Table shows the flow of the implemented gaze behavior.

Here, the data for the head rotation, gaze, and blink that were used in Behavior 6 is described. The graphs of head rotation, gaze, and blink that were used in Behavior 6 are shown in S3–S5 Figs.

The head rotation data consisted of three angles: yaw (shake), pitch (nod), and roll (tilt). The three-axis actuator is rotated according to these three angles. In Behavior 6, the pitch was 0°, which corresponded to the angle of eye contact with the local participant, combined with the gaze. Roll did not change from 0°, indicating the lack of head tilt. The yaw changed from -20°, the angle at which the eyes met local participant 2, to 10°, the angle at which the eyes met local participant 1, at 0.52 s after the behavior onset. Owing to the eye design, the absolute value of the angle of eye contact with local Participant 1 was smaller than that with local Participant 2.

The data for the change in yaw over 0.52 s were generated using the inverse tangent function, as shown in Eq. (S1).

$$\omega = 10^{\frac{2t-0.5}{0.25}}$$
$$yaw = \begin{cases} 5, \ if \ \omega = 1, \\ tan^{-1} \frac{2\zeta\omega}{1-\omega^2} - 20, if \ \omega \neq 1 \end{cases} \quad (S1)$$

The term t is the time from the beginning of the behavior, and $\zeta$ is a parameter that adjusts the slow-in/slow-out emphasis. The first term in the third row of Eq. (S1) is the inverted sign of the phase angle of the second-order delay element whose angular frequency is $\omega$ ($G(s) = \frac{1}{s^2+2\zeta s+1}$). Using Eq (S1), the emphasis of slow-in and slow-out can be adjusted while maintaining the time required for the yaw angle to change the constant.

The gaze data are represented as the position of the pupil in coordinates on the xy-plane with 0≤x≤100 and 0≤y≤100. The pupil image is moved to the position defined by the x–y coordinates in the eye display. The x-coordinate (horizontal coordinate) was displaced linearly in the direction of gaze shift and then returned slightly in the opposite direction in a time (0.31 s) shorter than the change in yaw angle. This indicates that the eyes moved earlier than the head rotated. The y coordinate (vertical coordinate) was displaced vertically downward with a blink. These eye movements imitated the behavioral features discovered by analyzing the eye-gaze data logs obtained from the avatar robot.

The blink data is an integer between 0 and 100, representing the degree of eye opening. At 0, the avatar robot's eyes are meditative, and at 100, its eyes are fully open. Blinks were performed at the beginning of the yaw change and immediately before the end of the yaw change. These were implemented as blinks before and after the head swing.

The data of the other behaviors were created by applying the data of Behavior 6.

The avatar robot changed its behavior at the start, during breathing, and at the end of the speech of the remote participant, as shown in S5 Table. We coded the start of speech as the time when the 16-bit PCM value of the remote participant's input audio exceeded a certain value. The breath-hold timing was coded as the time in which the 16-bit PCM value of the remote participant's input audio fell below a certain value during Behaviors 2–6. The end of the remote participant's speech was defined as the time at which the participant pressed the button.

**S2 Table. Flow of the implemented gaze behavior of the avatar robot.**

| Timing when gaze behavior takes place | Behavior number | Gaze behavior |
|---|---|---|
| Before a remote participant speaks. (In the controlled group, the avatar robot performed Behavior 1 while the researcher was speaking.) | 1 | Keep looking between two local participants. |
| When a remote participant begins to speak. | 2 | Direct its gaze at local participant 1. |
| After Behavior 2 or Behavior 6, until a remote participant takes a breath. | 3 | Keep looking at local participant 1. |
| When a remote participant takes a breath during Behavior 3. | 4 | Direct its gaze at local participant 2. |
| After Behavior 4, until a remote participant takes a breath. | 5 | Keep looking at local participant 2. |

| When a remote participant takes a breath during Behavior 5. | 6 | Direct its gaze at local participant 1. |
| At the end of a remote participant's speech. If the avatar robot is in Behavior 3, Behavior 7 starts; if the avatar robot is in Behavior 5, Behavior 8 starts. | 7,8 | Direct its gaze between two local participants. |

48

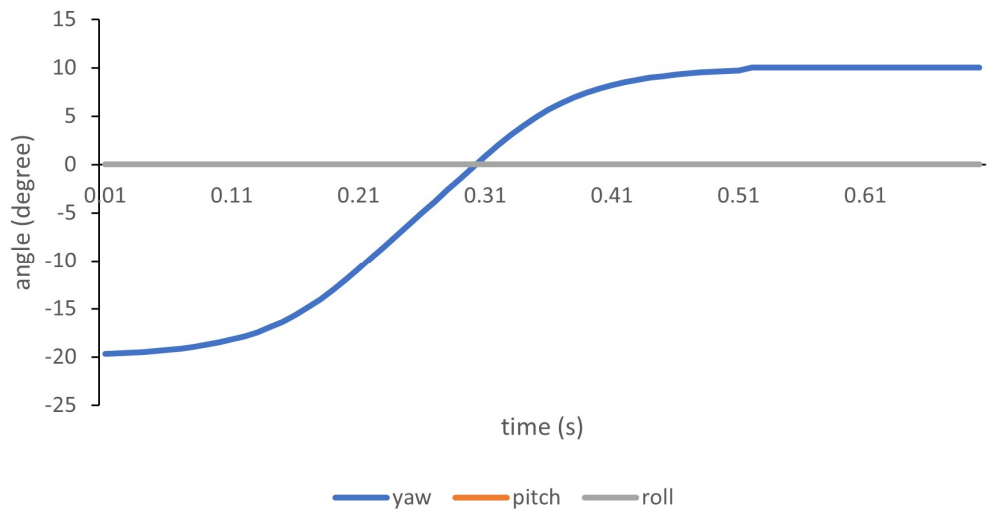

49
50  **S3 Fig. Head rotation of the avatar robot in Behavior 6.** Using these data as input, we
51  controlled the head rotation of the avatar robot.
52

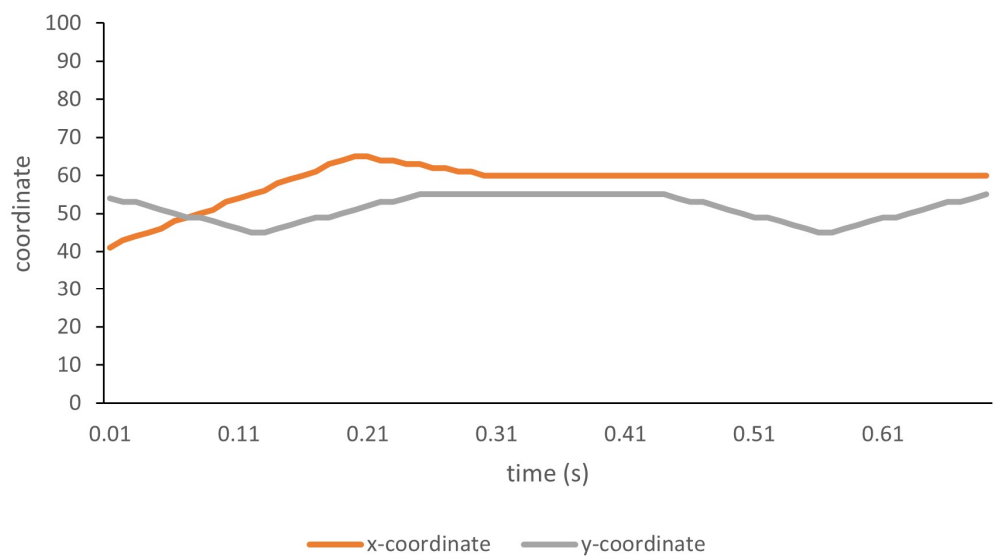

53
54  **S4 Fig. Gaze of the avatar robot in Behavior 6.** The gaze data are the pupil positions expressed as coordinates in the xy-plane with 0≤x≤100,　0≤y≤100. The pupil positions of the two eyes are made to be the same.

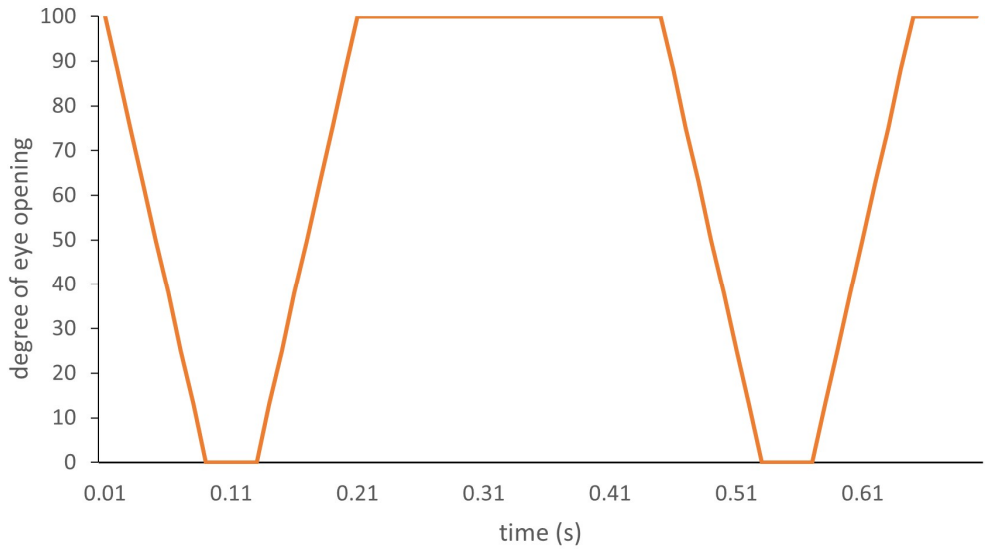

**S5 Fig. Blink of the avatar robot in Behavior 6.** The blink data were integers from 0 to 100, representing the degree of eye opening; at 0, the avatar robot's eyes were closed, and at 100, they were fully open.